\documentclass[aps,jcp,twocolumn,amsmath,amssymb,amsfonts,floatfix,superscriptaddress]{revtex4}

\usepackage{graphicx}
\usepackage{lscape}
\usepackage{float}
\usepackage{psfrag}
\usepackage{algpseudocode}
\usepackage{appendix}
\usepackage{caption}
\usepackage{subfig}
\usepackage{amssymb}
\usepackage{threeparttable}

\begin{document}

\title{A perturbative approximation to DFT/MRCI: DFT/MRCI(2)}

\author{Simon P. Neville}
\affiliation{National Research Council Canada, 100 Sussex Drive,
  Ottawa, Ontario K1A 0R6, Canada}

\author{Michael S. Schuurman}
\affiliation{National Research Council Canada, 100 Sussex Drive,
  Ottawa, Ontario K1A 0R6, Canada}
\affiliation{Department of Chemistry and Biomolecular Sciences,
  University of Ottawa, 10 Marie Curie, Ottawa, Ontario, K1N 6N5,
  Canada}

%----------------------------------------------------------------------
% Abstract
%----------------------------------------------------------------------
\begin{abstract}
  We introduce a perturbative approximation to the combined density
  functional theory and multireference configuration interaction
  (DFT/MRCI) method. The method, termed DFT/MRCI(2), results from the
  application of quasi-degenerate perturbation theory and the
  Epstein-Nesbet partitioning of the DFT/MRCI Hamiltonian matrix. This
  results in the replacement of the diagonalization of the large
  DFT/MRCI Hamiltonian with that of a small effective Hamiltonian, and
  affords orders of magnitude savings in terms of computational
  cost. Moreover, the DFT/MRCI(2) approximation is found to be of
  excellent accuracy, furnishing excitation energies with a root mean
  squared deviation from the DFT/MRCI values of less than 0.03 eV for
  an extensive test set of organic molecules.
\end{abstract}

\maketitle

%----------------------------------------------------------------------
% Introduction
%----------------------------------------------------------------------
\section{Introduction}
First formulated by Grimme and Waletzke\cite{grimme_dft-mrci}, the
combined density functional theory and multireference configuration
interaction (DFT/MRCI) method fills a unique niche in the landscape of
excited-state electronic structure theory. In particular, DFT/MRCI
possesses the following desirable characteristics: (i) the possibility
of a ``black box'' implementation; (ii) high accuracy in vertical
excitation energies; (iii) low computational cost, rendering it
applicable to large molecules, and; (iv) an ability to describe a
large range of electronic states, including those of multireference,
Rydberg, doubly-excited and charge transfer character.

Originally conceived as a method for the description of singlet and
triplet states of organic molecules\cite{grimme_dft-mrci}, recent
years have seen the development of DFT/MRCI Hamiltonians that are
multiplicity agnostic\cite{lyskov_dftmrci_redesign, heil_dftmrci_2017,
  marian_dftmrci_review}, and tailored to the description of
transition metal complexes\cite{heil_dftmrci_transition_metals}. It
has also been recently demonstrated that DFT/MRCI may be successfully
applied to the calculation of core-excited states via the application
of the core-valence separation
approximation\cite{seidu_cvsdftmrci}. In addition to excitation
energies, a plethora of properties may be reliably computed using
DFT/MRCI, including spin-orbit coupling\cite{kelinschmidt_spock_ci,
  kleinschmidt_spin_orbit_coupling_constants}, diabatic potentials and
couplings\cite{ours_dftmrci_pbdd}, and numerical derivative
couplings\cite{bracker_dftmrci_derivative_coupling}.

Recently, a new DFT/MRCI pruning algorithm was
introduced\cite{neville_p-dftmrci}, in which deadwood configurations
are removed based on an estimate of their contribution to the wave
functions of interest made using Rayleigh Schr\"{o}dinger perturbation
theory (RSPT) in conjunction with the Epstein-Nesbet Hamiltonian
partitioning\cite{epstein26, nesbet55}. The resulting method, termed
p-DFT/MRCI, was shown to retain the accuracy of the original DFT/MRCI
method, but with configuration spaces reduced in size by up to two
orders of magnitude. This result suggests that perturbative
approximations based on the Epstein-Nesbet Hamiltonian partitioning
are well suited to pairing with the DFT/MRCI. We here explore taking
this approach to its logical conclusion: the complete replacement of
the expensive diagonalization of the DFT/MRCI Hamiltonian matrix with
a perturbative approximation of its eigenpairs.

The approach pursued here is to use quasi-degenerate perturbation
theory (QDPT) to replace the DFT/MRCI Hamiltonian with a small
effective Hamiltonian that may be fully diagonalized to yield
approximations to the DFT/MRCI states and energies. In particular, we
explore the combination of generalized van Vleck perturbation theory
(GVVPT)\cite{kirtman_gvvpt, hirschfelder78, certain70,
  certain70_erratum} with the Epstein-Nesbet Hamiltonian
partitioning. The resulting method, which we term DFT/MRCI(2), is
shown to yield excellent approximations to the DFT/MRCI excitation
energies, exhibiting a root mean square deviation of the order of
$10^{-2}$ eV for a comprehensive test set. Moreover, the cost of
building and diagonalizing the DFT/MRCI(2) effective Hamiltonian
matrix is dramatically reduced in comparison to the iterative
diagonalization of the full DFT/MRCI Hamiltonian and affords
computational savings expected to reach three orders of magnitude for
large molecules.

The rest of the paper is structured as follows. In
Section~\ref{sec:dftmrci}, we provide a brief overview of the DFT/MRCI
method. In Section~\ref{sec:dftmrci(2)}, we give the details of the
DFT/MRCI(2) approximation and working equations. A set of benchmarking
and example calculations are presented in Section~\ref{sec:results},
followed by concluding remarks and outlooks in
Section~\ref{sec:conclusions}.

%----------------------------------------------------------------------
% DFT/MRCI
%----------------------------------------------------------------------
\section{DFT/MRCI}\label{sec:dftmrci}

\subsection{The DFT/MRCI wave function and Hamiltonian matrix}

In common with all MRCI approaches, the DFT/MRCI wave function
{\it ansatz} may be written as

\begin{equation}
  | \Psi_{I} \rangle = \sum_{\Omega \in \mathcal{R}} C_{\Omega}^{(I)}
  |\Omega\rangle + \sum_{\Omega \in \mathcal{F}} C_{\Omega}^{(I)}
  |\Omega\rangle,
\end{equation}

\noindent
where, $|\Omega\rangle$ denote spin-adapted configuration state
functions (CSFs). The total CSF space is partitioned into a reference
space $\mathcal{R}$ and the first-order interacting space (FOIS)
$\mathcal{F}$, obtained by the application of one- and two-electron
excitation operators to $\mathcal{R}$. Here, the comparatively small
reference space is chosen to yield a qualitatively correct description
of the states $| \Psi_{I} \rangle$ of interest, capturing the
pertinent static electron correlation. The FOIS CSFs account for the
remaining dynamic electron correlation. The FOIS, in contrast to the
reference space, is generally large in size, and the diagonalization
of the electronic Hamiltonian projected onto the space spanned by
$\mathcal{R} \cup \mathcal{F}$ quickly becomes an extremely demanding
computational task.

The DFT/MRCI method seeks to circumvent this issue through the
introduction of DFT-specific on-diagonal Hamiltonian matrix
corrections to recover the majority of the dynamic correlation that
would otherwise be captured by the interaction of the reference space
and FOIS CSFs. Let each CSF $|\Omega\rangle$ be specified by a spatial
orbital occupation $\textbf{w}$ and a spin-coupling pattern $\omega$:
$|\Omega\rangle = |\textbf{w} \omega\rangle$. Then, the DFT-specific
corrections take the form of the following modifications of the
on-diagonal Hamiltonian matrix elements\cite{grimme_dft-mrci}:

\begin{equation}\label{eq:dftmrci_ondiag}
  \begin{aligned}
    \left\langle \textbf{w} \omega \middle| \hat{H}^{DFT} - E_{DFT}
    \middle| \textbf{w} \omega \right\rangle &= \left\langle
    \textbf{w} \omega \middle| \hat{H} - E_{SCF} \middle| \textbf{w}
    \omega \right\rangle \\
    &+ \sum_{p} \Delta \text{w}_{p} \left( \epsilon_{p}^{KS} - F_{pp}
    \right) \\
    &+ \Delta E_{x} + \Delta E_{c}.
  \end{aligned}
\end{equation}

\noindent
Here, $\Delta \text{w}_{p} = \text{w}_{p} - \overline{\text{w}}_{p}$
denotes the difference of the occupation of the $p$th spatial orbital
relative to a base, or anchor, occupation $\overline{\textbf{w}}$,
chosen as the Hartree-Fock (HF) occupation. $\epsilon_{p}^{KS}$ and
$F_{pp}$ denote, respectively, the Kohn-Sham (KS) orbital energies and
on-diagonal elements of the Fock operator in the KS orbital
basis. Finally, $\Delta E_{x}$ and $\Delta E_{c}$ are Coulomb and
exchange corrections, the exact form of which varies with the
different DFT/MRCI parameterizations\cite{grimme_dft-mrci,
  lyskov_dftmrci_redesign, heil_dftmrci_2017,
  heil_dftmrci_transition_metals}.

Underlying these semi-empirical corrections is the recognition that
differences between diagonal Fock matrix elements contribute, in a
zeroth-order manner, to the on-diagonal Hamiltonian matrix elements
(see, e.g., the work of Segal, Wetmore and Wolf\cite{wetmore_1975,
  segal_1978}). The differences between KS orbital energies are
typically closer to ground-to-excited-state excitation energies than
the corresponding differences in Fock matrix elements. Building these
into the on-diagonal Hamiltonian matrix elements, and appropriately
down-scaling the Coulomb and exchange terms that appear alongside, one
may effectively incorporate a large amount of dynamic electron
correlation that would otherwise be accounted for by the coupling of
the reference and FOIS CSFs. To avoid a double counting of dynamic
correlation, the off-diagonal Hamiltonian matrix elements must be
appropriately adjusted. This is achieved through the introduction of a
damping of the off-diagonal elements that is dependent on the
energetic separation of the bra and ket CSFs:

\begin{equation}\label{eq:dftmrci_offdiag}
  \begin{aligned}
    \left\langle \textbf{w} \omega \middle| \hat{H}^{DFT} - E_{DFT}
    \middle| \textbf{w}' \omega' \right\rangle &= \left\langle \Omega
    \middle| \hat{H} - E_{SCF} \middle| \Omega' \right\rangle \\
    &\times D(\Delta E_{\Omega\Omega'}),
  \end{aligned}
\end{equation}

\noindent
where

\begin{equation}
  \Delta E_{\Omega\Omega'} = \frac{1}{n_{\omega}}
  \sum_{\omega}^{n_{\omega}}
  H_{\Omega,\Omega}^{DFT} - \frac{1}{n_{\omega'}}
  \sum_{\omega'}^{n_{\omega'}}
  H_{\Omega', \Omega'}^{DFT}
\end{equation}

\noindent
denotes the spin coupling-averaged difference between the on-diagonal
matrix elements corresponding to the spatial occupations $\textbf{w}$
and $\textbf{w}'$, which generate $n_{\omega}$ and $n_{\omega'}$ CSFs,
respectively. The damping function $D(\Delta E)$ is chosen to decay
rapidly with increasing $\Delta E$: in practice chosen as either an
exponential\cite{grimme_dft-mrci, heil_dftmrci_transition_metals} or
inverse arctangent\cite{lyskov_dftmrci_redesign, heil_dftmrci_2017}
function. In this way, the coupling of energetically distant reference
and FOIS CSFs is damped to near-zero, thereby avoiding to a large
extent a double counting of dynamic correlation effects.

This decoupling of a large part of the FOIS from the reference space
means that most of the FOIS CSFs are no longer required. These are
identified \textit{a priori} using a simple orbital energy-based
selection criterion\cite{grimme_dft-mrci}, which proceeds as follows.
For each FOIS configuration $\textbf{w}$, the quantity

\begin{equation}\label{eq:esel}
  d_{\textbf{w}}= \sum_{p} \Delta \text{w}_{p} \epsilon_{p}^{KS} -
  \delta E_{sel}
\end{equation}

\noindent
is computed, where $\delta E_{sel}$ is a parameter with a value
conventionally chosen as either 1.0 or 0.8 E$_{\text{h}}$. If
$d_{\textbf{w}}$ is less than the highest reference space eigenvalue
of interest, then all the CSFs generated from the configuration
$\textbf{w}$ are selected for inclusion, else they are discarded. This
configuration selection step results in a massive reduction of the
size of the CSF basis, typically by many orders of magnitude, and
results in huge speedups relative to an \textit{ab initio} MRCI
calculation.

\subsection{Reference space selection and refinement}\label{sec:autoras}
One appealing property of DFT/MRCI is that, being an individually
selecting CI method, arbitrary reference spaces may be used, imbuing
the method with great flexibility. Moreover, a DFT/MRCI calculation is
typically performed in the following iterative fashion. First, a
standard DFT/MRCI calculation is performed using an initial, guess
reference space $\mathcal{R}_{0}$. The resulting eigenvectors are then
analyzed and a new, improved reference space $\mathcal{R}_{1}$ is
constructed which contains all the configurations that generate one or
more of the CSFs contributing significantly to the states of
interest. Another DFT/MRCI calculation is then performed using the
refined reference space, and this procedure is repeated until the
reference space is converged.

This reference space refinement procedure may be automated, and all
that is left is to specify the initial reference space
$\mathcal{R}_{0}$. Recently, a fully automated algorithm for the
construction of $\mathcal{R}_{0}$ was
introduced\cite{neville_p-dftmrci}, in which a preliminary approximate
combined DFT and CI singles (DFT/CIS)\cite{grimme_dftcis} calculation
is used to identify a subset of orbitals from which compact, but
qualitatively correct, restricted active space CI
(RASCI)\cite{olsen1988} reference spaces may be constructed. The
result is a completely black box calculation, requiring only the
number of states to be specified by the user. Moreover, the ability to
automatically and reliably construct initial reference spaces that
yield good zeroth-order descriptions of the states of interest is of
utmost importance if perturbative approximations are to be introduced,
as is the focus this paper.

%----------------------------------------------------------------------
% DFT/MRCI(2)
%----------------------------------------------------------------------
\section{DFT/MRCI(2)}\label{sec:dftmrci(2)}
To arrive at a robust and easily implemented perturbative
approximation to the DFT/MRCI method, we consider the application of
the Epstein-Nesbet Hamiltonian partitioning\cite{epstein26, nesbet55}
in conjunction with QDPT\cite{brandow_1967, lowdin_1971,
  lindgren_1974, klein_1974, kvasnicka_1977, hose_1979, shavitt80}. In
particular, we consider use of the GVVPT variant of
QDPT\cite{kirtman_gvvpt}. This combination, when applied to the
DFT/MRCI Hamiltonian, results in the DFT/MRCI(2) approximation.

\subsection{Generalized van Vleck perturbation theory}
To keep the present discussion self-contained, we will here give a
brief overview of the generalized van Vleck variant of QDPT, focusing
on the aspects pertinent to the development of the DFT/MRCI(2)
method. The starting point is the partitioning of the total
Hamiltonian $\hat{H}$ as the sum of a zeroth-order operator
$\hat{H}_{0}$ and a perturbation $\hat{V}$:

\begin{equation}\label{eq:hpart}
  \hat{H} = \hat{H}_{0} + \hat{V}.
\end{equation}

\noindent
Let $| \Psi_{k}^{(0)} \rangle$ denote the eigenfunctions of the
zeroth-order Hamiltonian $\hat{H}_{0}$, with associated eigenvalues
$E_{k}^{(0)}$,

\begin{equation}
  \hat{H}_{0} | \Psi_{k}^{(0)} \rangle = E_{k}^{(0)} | \Psi_{k}^{(0)}
  \rangle.
\end{equation}

\noindent
In the GVVPT scheme, the set $\{ |\Psi_{k}^{(0)}\rangle \}$ is
subdivided into a ``model space'' $\mathcal{M} = \{|\phi_{i}\rangle\}$
and an ``external space'' $\mathcal{E} = \{|\chi_{a}\rangle\}$. The
model space $\mathcal{M}$ is to be equated with with the zeroth-order
states $|\Psi_{k}^{(0)} \rangle$ for which perturbative corrections
are sought, and the external space $\mathcal{E}$ with their orthogonal
complement.

In common with all QDPT variants, in GVVPT a similarity transformation
$\boldsymbol{U}$ is sought that block diagonalizes the Hamiltonian
such that the model and external spaces are decoupled:

\begin{equation}
  \boldsymbol{U}^{-1} \boldsymbol{H} \boldsymbol{U} =
  \boldsymbol{\mathcal{H}}=
  \begin{bmatrix}
    \boldsymbol{\mathcal{H}}_{\mathcal{M}\mathcal{M}} & \boldsymbol{0} \\
    \boldsymbol{0} & \boldsymbol{\mathcal{H}}_{\mathcal{E}\mathcal{E}}
  \end{bmatrix}.
\end{equation}

\noindent
Diagonalization of the model space block,
$\boldsymbol{\mathcal{H}}_{\mathcal{M}\mathcal{M}}$, of the effective
Hamiltonian $\boldsymbol{\mathcal{H}}$ then yields the energies of the
states of interest.

%Let $\{ |\tilde{\phi}_{i}\rangle \}$ and $\{ |\tilde{\chi}_{a}\rangle
%\}$ denote the sets of perturbed model and external states,
%respectively, that result from the block diagonalization,
%
%\begin{equation}
%  |\tilde{\phi}_{i}\rangle = \sum_{l} U_{li} |\Psi_{l}^{(0)}\rangle, 
%\end{equation}
%
%\begin{equation}
%  |\tilde{\chi}_{a}\rangle = \sum_{l} U_{la} |\Psi_{l}^{(0)}\rangle.
%\end{equation}

Let $\{ |\tilde{\phi}_{i}\rangle \}$ and $\{ |\tilde{\chi}_{a}\rangle
\}$ denote the sets of perturbed model and external states,
respectively, that result from the block diagonalization
transformation $\boldsymbol{U}$. To proceed, these are expanded in a
perturbation series,

\begin{equation}
|\tilde{\phi}_{i}\rangle = |\phi_{i}\rangle + |\phi_{i}^{(1)}\rangle
  + |\phi_{i}^{(2)}\rangle + \cdots,
\end{equation}

\begin{equation}
  |\tilde{\chi}_{a}\rangle = |\chi_{a}\rangle + |\chi_{a}^{(1)}\rangle
  + |\chi_{a}^{(2)}\rangle + \cdots.
\end{equation}

\noindent
In the GVVPT method, perturbative approximations for the model space
block, $\boldsymbol{\mathcal{H}}_{\mathcal{M}\mathcal{M}}$ of the
effective Hamiltonian and the perturbed model states $\{
|\tilde{\phi}_{i}\rangle \}$ are determined by the enforcement of the
following two conditions: (i) that the effective Hamiltonian is block
diagonal in the perturbed basis, and; (ii) that the perturbed basis is
orthogonal within the model space and between the model and external
spaces. This results in the following form of the second-order
effective Hamiltonian within the model space\cite{kirtman_gvvpt}:

\begin{equation}\label{eq:gvvpt2_ham}
  \begin{aligned}
    \mathcal{H}_{ij}^{[2]} &= \langle \phi_{i} | \hat{H}_{0} |
    \phi_{j} \rangle + \langle \phi_{i} | \hat{V} | \phi_{j} \rangle\\
    &+ \frac{1}{2} \left[ \langle \phi_{i} | \hat{V} \hat{R}_{j}
      \hat{V} | \phi_{j} \rangle + \langle \phi_{i} | \hat{V}
      \hat{R}_{i} \hat{V} | \phi_{j} \rangle \right],
  \end{aligned}
\end{equation}

\noindent
where the resolvant operator $\hat{R}_{i}$ is defined as

\begin{equation}
  \hat{R}_{i} = \sum_{a} |\chi_{a}\rangle \left( E_{i}^{(0)} -
  E_{a}^{(0)} \right)^{-1} \langle \chi_{a} |
\end{equation}

\noindent
The first-order perturbed model states are given by

\begin{equation}
  |\tilde{\phi}_{i}^{[1]}\rangle = |\phi_{i}\rangle +
  \hat{R}_{i}\hat{V} |\phi_{i}\rangle.
\end{equation}

Diagonalization of the model space block of
$\boldsymbol{\mathcal{H}}^{[2]}$ yields the second-order GVVPT
(GVVPT2) approximations, $E_{I}^{[2]}$, of the energies of the states
of interest,

\begin{equation}
  \boldsymbol{E}^{[2]} = \text{diag}\left( E_{1}^{[2]},
  E_{2}^{[2]}, \dots, E_{m}^{[2]} \right) = \boldsymbol{X}^{T}
  \boldsymbol{\mathcal{H}}_{\mathcal{M}\mathcal{M}}^{[2]}
  \boldsymbol{X}
\end{equation}

\noindent
Finally, first-order corrected wave functions $|\Psi_{I}^{[1]}\rangle$
may be computed using the eigenvectors $\boldsymbol{X}$ and
first-order perturbed model states $|\tilde{\phi}_{i}^{[1]}\rangle$:

\begin{equation}
  |\Psi_{I}\rangle \approx |\Psi_{I}^{[1]}\rangle = \sum_{i} X_{iI}
  |\tilde{\phi}_{i}^{[1]}\rangle.
\end{equation}

\subsection{Epstein-Nesbet Hamiltonian partitioning}
To arrive at working equations, the form of the Hamiltonian
partitioning (i.e., $\hat{H}_{0}$ and $\hat{V}$) needs to be
specified. We adopt the Epstein-Nesbet partitioning, in which the
total set of CSFs $\{ \Omega \}$ is partitioned into two subspaces,
termed the $\mathcal{P}$ and $\mathcal{Q}$ spaces. The zeroth-order
Hamiltonian is taken to be block diagonal between the $\mathcal{P}$
and $\mathcal{Q}$ spaces and diagonal within the $\mathcal{Q}$ space:

\begin{equation}\label{eq:enh0}
  \hat{H}_{0} = \sum_{\Omega, \Omega' \in \mathcal{P}} | \Omega
  \rangle \langle \Omega | \hat{H} | \Omega' \rangle \langle \Omega' |
  + \sum_{\Omega \in \mathcal{Q}} | \Omega \rangle \langle \Omega |
  \hat{H} | \Omega \rangle \langle \Omega |.
\end{equation}

\noindent
The perturbation accounts for the missing coupling between the
$\mathcal{P}$ and $\mathcal{Q}$ spaces and that within the
$\mathcal{Q}$ space:

\begin{equation}\label{eq:env}
  \begin{aligned}
    \hat{V} &= \left\{ \sum_{\substack{\Omega \in \mathcal{P}
        \\ \Omega' \in \mathcal{Q}}} | \Omega \rangle \langle \Omega |
    \hat{H} | \Omega' \rangle \langle \Omega' | + h.c. \right\} \\
    &+ \sum_{\substack{\Omega, \Omega' \in \mathcal{Q} \\ \Omega \ne
        \Omega'}} | \Omega \rangle \langle \Omega | \hat{H} | \Omega'
    \rangle \langle \Omega' |
  \end{aligned}
\end{equation}

\noindent
Within the context of a DFT/MRCI calculation, the natural choice is to
identify the $\mathcal{P}$ space with the reference space
($\mathcal{P}=\mathcal{R}$) and the $\mathcal{Q}$ with the FOIS CSFs
that have survived the configuration selection criterion
($\mathcal{Q}\subset\mathcal{F}$), and this shall be assumed in the
following.

\subsection{DFT/MRCI(2) working equations}
We now consider the combination of GVVPT2 with the Epstein-Nesbet
Hamiltonian partitioning and, in particular, the application of these
to the DFT/MRCI Hamiltonian matrix. In order to calculate the energies
of the $N$ states, we choose that the model space to be spanned by the
$N$ lowest-lying reference space eigenfunctions of interest,
$|\Psi_{I}^{(0)}\rangle$, with corresponding energies $E_{I}^{(0)}$,
plus a small set of additional `buffer' states $\{
|\Psi_{I}^{(0)}\rangle | I=N+1,\dots,N+N_{buf} \}$. Substitution of
Equations~\ref{eq:enh0} and \ref{eq:env} into
Equation~\ref{eq:gvvpt2_ham}, and replacing the \textit{ab initio}
Hamiltonian $\hat{H}$ with the DFT/MRCI Hamiltonian $\hat{H}^{DFT}$,
we obtain the following expression for the DFT/MRCI(2) effective
Hamiltonian:

\begin{equation}\label{eq:dftmrci2_ham}
  \left[\boldsymbol{\mathcal{H}}_{DFT}^{[2]}\right]_{IJ} = \delta_{IJ}
  E_{I}^{(0)} + \frac{1}{2} \sum_{i=I,J} \sum_{\Omega \in \mathcal{Q}}
  \frac{B_{\Omega I} B_{\Omega J}}{E_{i}^{(0)} - E_{\Omega}^{(0)}},
\end{equation}

\noindent
where

\begin{equation}
  B_{\Omega I} = \langle \Omega | \hat{H}^{DFT} | \Psi_{I}^{(0)}
  \rangle,
\end{equation}

\begin{equation}
  E_{\Omega}^{(0)} = \langle \Omega | \hat{H}^{DFT} | \Omega \rangle,
\end{equation}

\noindent
and real CSFs have been assumed.

Similarly, the DFT/MRCI(2) first-order corrected eigenfunctions are
given by

\begin{equation}\label{eq:dftmrci2_psi}
  |\Psi_{I}^{[1]}\rangle = \sum_{J} X_{JI} \left[
    |\Psi_{J}^{(0)}\rangle + \sum_{\Omega \in \mathcal{Q}} \left(
    \frac{B_{\Omega J}}{E_{J}^{(0)} - E_{\Omega}^{(0)}} \right)
    |\Omega\rangle \right],
\end{equation}

\noindent
where $\boldsymbol{X}$ is the matrix of eigenvectors of
$\boldsymbol{\mathcal{H}}_{DFT}^{[2]}$.

We note that all the matrix elements required to evaluate
Equations~\ref{eq:dftmrci2_ham} and \ref{eq:dftmrci2_psi} are also
needed in the calculation of the DFT/MRCI Hamiltonian matrix. Thus, it
is a somewhat trivial task to modify an existing DFT/MRCI code to
compute the DFT/MRCI(2) effective Hamiltonian and first-order
corrected eigenfunctions. We emphasize here that the calculation of
these quantities requires significantly less computational effort than
the iterative diagonalization of the DFT/MRCI Hamiltonian matrix.

\subsection{Dynamical reference space refinement}\label{sec:dyn_ref_sel}
The refinement of the DFT/MRCI(2) reference space proceeds in an
analogous manner to that of a DFT/MRCI calculation except that the
DFT/MRCI wave functions are replaced with their first-order corrected
approximations $| \Psi_{I}^{[1]} \rangle$. The automated,
DFT/CIS-based initial reference space generation algorithm described
in Section~\ref{sec:autoras} is used in our implementation of
DFT/MRCI(2). This generally results in initial reference spaces with
excellent support of the states of interest. However, in practice, we
have encountered a small number of situations in which the thus
generated initial reference space is only of moderate quality, as
measured by the norm of the projection of the first-order corrected
eigenfunctions onto the $\mathcal{Q}$ space,

\begin{equation}
  Q_{I} = \left|\left| \sum_{\Omega \in \mathcal{Q}} |\Omega\rangle
  \langle \Omega| \Psi_{I}^{[1]} \rangle \right|\right|.
\end{equation}

\noindent
When $Q_{I}$ is large, the first-order corrected eigenfunctions may
only be qualitatively correct. In such a case, a more relaxed
configuration selection threshold may be beneficial in the reference
space refinement step. We have thus implemented the following
state-specific, dynamical configuration selection threshold:

\begin{equation}\label{eq:dynsel}
  \epsilon(Q_{I}) = \max\left(\epsilon_{min},
  \frac{\epsilon_{max}}{\cosh^{2}\left(\mu Q_{I} \right)} \right).
\end{equation}

\begin{figure}
  \begin{center}
    \includegraphics[width=0.45\textwidth,angle=0]{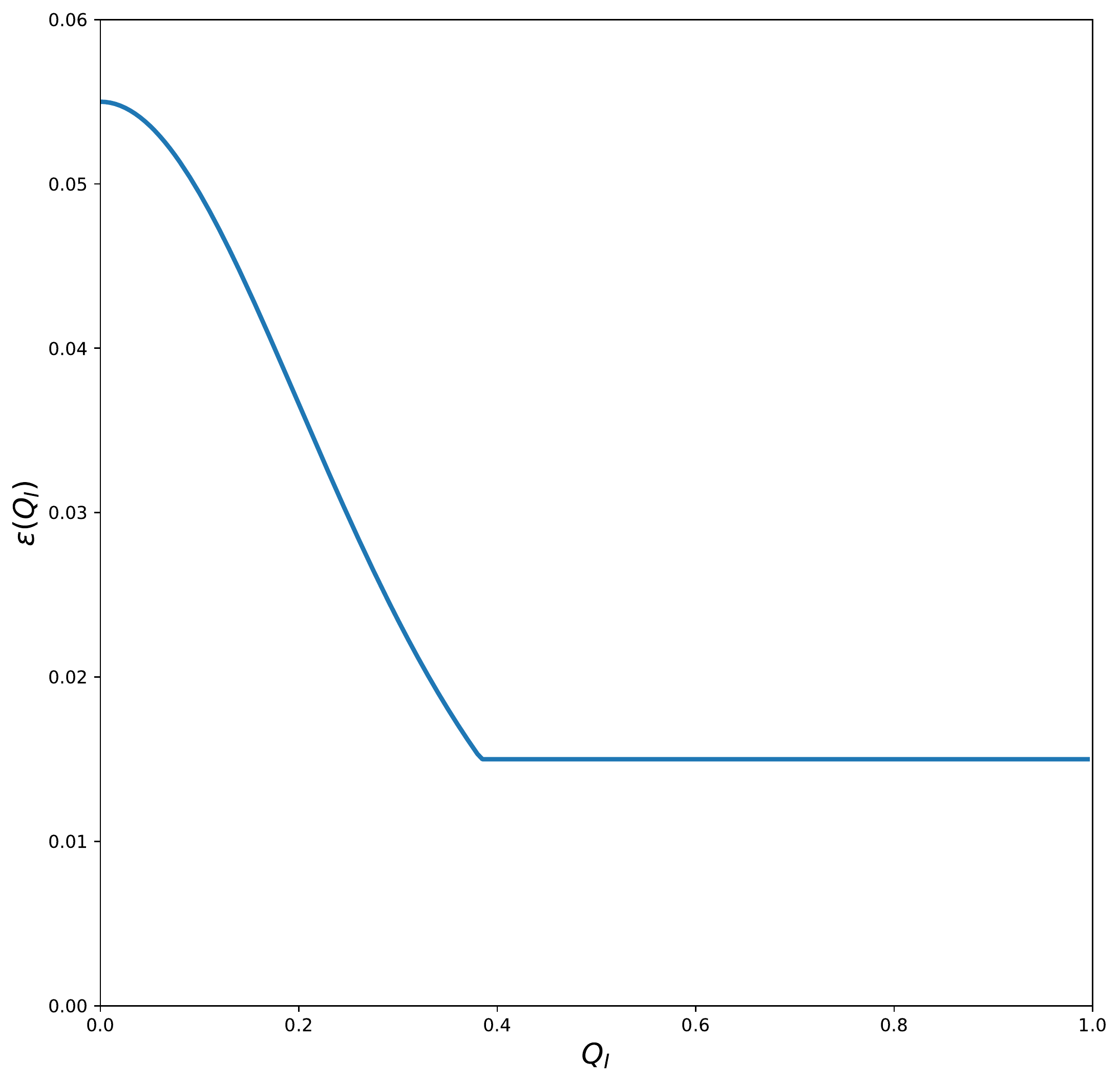}
    \caption{Dynamic configuration selection threshold function
      $\epsilon(Q_{I})$ used in all DFT/MRCI(2) calculations for
      parameter values $\epsilon_{min}=0.015$, $\epsilon_{max}=0.055$,
      and $\mu=3.3$. See Equation \ref{eq:dynsel} for the definition
      of this function.}
    \label{fig:dynsel}
  \end{center}
\end{figure}

\noindent
If the CSF $|\Omega\rangle$ has a coefficient for state $I$ with an
absolute value greater than $\epsilon(Q_{I})$, then the corresponding
configuration is chosen for inclusion in the refined reference
space. In this work, empirical parameter values of
$\epsilon_{min}=0.015$, $\epsilon_{max}=0.055$, and $\mu=3.3$ have
been used. For reference, the dynamic selection threshold
$\epsilon(Q_{I})$ is shown plotted in Figure~\ref{fig:dynsel} for
these values.

\subsection{Intruder states}\label{sec:isa}
In practice, there may exist $\mathcal{Q}$ space CSFs $|\Omega\rangle$
that are near degenerate with one of the reference space states
$|\Psi_{I}^{(0)}\rangle$ of interest. This is most likely to occur in
the first iteration of the reference space refinement in which a
guess/unrefined reference space $\mathcal{R}_{0}$ is employed. For a
given pair of states $I$ and $J$, we will obtain effective Hamiltonian
matrix elements $\left|
\left[\boldsymbol{\mathcal{H}}_{DFT}^{[2]}\right]_{IJ} \right|
\rightarrow \infty$ if there exists a $\mathcal{Q}$ space CSF
$|\Omega\rangle$ satisfying

\begin{equation}
  \left| E_{i}^{(0)} - E_{\Omega}^{(0)} \right| \rightarrow
  0, \hspace{0.5cm} i=I \text{ or } i=J.
\end{equation}

\noindent
Such a CSF is referred to as an ``intruder state''.

To ameliorate this problem, we adopt the intruder state avoidance
(ISA) technique of Witek \textit{et al.}\cite{witek_isa}. Here, the
energy denominators appearing in Equations~\ref{eq:dftmrci2_ham} and
\ref{eq:dftmrci2_psi} are subjected to the replacement

\begin{equation}
  \left( E_{I}^{(0)} - E_{\Omega}^{(0)} \right)^{-1} \rightarrow
  \left( E_{I}^{(0)} - E_{\Omega}^{(0)} + \Delta_{\Omega I}
  \right)^{-1},
\end{equation}

\noindent
where

\begin{equation}
  \Delta_{\Omega I} = \frac{b}{E_{I}^{(0)} - E_{\Omega}^{(0)}},
\end{equation}

\noindent
with $b \in \mathbb{R}$ being a free parameter. The shift
$\Delta_{\Omega I}$ satisfies

\begin{equation}
  \left| \Delta_{\Omega I} \right| \rightarrow \infty
  \hspace{0.5cm} \text{as} \hspace{0.5cm} \left| E_{I}^{(0)} -
  E_{\Omega}^{(0)} \right| \rightarrow 0.
\end{equation}

\noindent
Thus, the contributions from any intruder states are damped out, and
the appearance of (near) singularities in the DFT/MRCI(2) working
equations is avoided.

Care must be taken in the application of the ISA technique within the
framework of a DFT/MRCI(2) calculation due to the iterative reference
space refinement procedure. Here, the initial/guess reference space
being updated to include all configurations contributing significantly
to the DFT/MRCI(2) first-order corrected eigenstates
$|\Psi_{I}^{[1]}\rangle$. However, the coefficients for any intruder
states will be damped to near zero when using the ISA scheme,
resulting in their exclusion from the refined reference space. To
remedy this, the terms

\begin{equation}
  C_{\Omega I} = B_{\Omega I} \left( E_{I}^{(0)} - E_{\Omega}^{(0)}
  \right)^{-1}
\end{equation}

\noindent
and

\begin{equation}
  \overline{C}_{\Omega I} = B_{\Omega I} \left( E_{I}^{(0)} -
  E_{\Omega}^{(0)} + \Delta_{\Omega I} \right)^{-1}
\end{equation}

\noindent
are computed during the course of the calculation of the effective
Hamiltonian. Let $\epsilon_{max}$ denote the maximum configuration
selection threshold used in the reference space refinement step (see
Section~\ref{sec:dyn_ref_sel}). If $|C_{\Omega I}| > \epsilon_{max}$
and $|\overline{C}_{\Omega I}| < \epsilon_{max}$ for any state $I$,
then the configuration generating the CSF $|\Omega\rangle$ is flagged
for explicit inclusion in the refined reference space. In this way,
any intruder states may be reliably identified and removed from the
$\mathcal{Q}$ space.

%----------------------------------------------------------------------
% Results
%----------------------------------------------------------------------
\section{Implementation and Results}\label{sec:results}
The DFT/MRCI(2) method was implemented in the General Reference
Configuration Interaction (GRaCI) program\cite{graci}. The required KS
MOs and integrals were computed using the PySCF package\cite{sun_2018,
  sun_2020}. Two-electron integrals were calculated using the the
density fitting approximation\cite{whitten_1973, feyereisen_1993,
  vahtras_1993}. In all calculations, the original DFT/MRCI
Hamiltonian of Grimme\cite{grimme_dft-mrci} was used, which is
parameterised for use with the BHLYP functional. An energy-based
configuration selection threshold of $\delta E_{sel} = 1.0$
E$_{\text{h}}$ was used in both the DFT/MRCI and DFT/MRCI(2)
calculations. In the construction of the DFT/MRCI(2) effective
Hamiltonian, $N_{buf}=10$ buffer states per irreducible representation
(irrep) were used along with an ISA shift parameter of
$b=5\times10^{-3}$ $E_{\text{h}}$.

\subsection{Benchmarking}\label{sec:vee}

\begin{figure}
  \begin{center}
    \includegraphics[width=0.45\textwidth,angle=0]{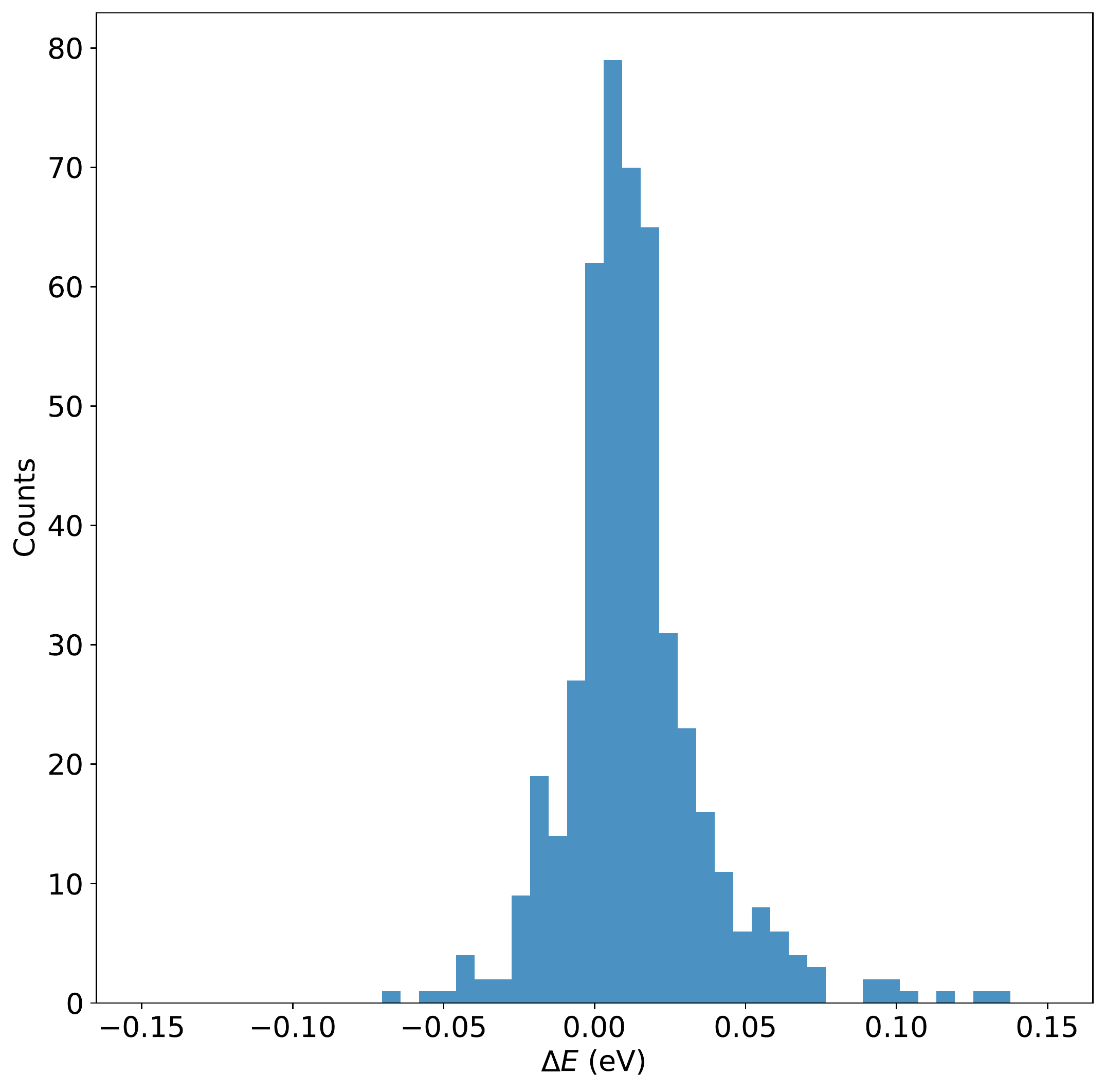}
    \caption{Differences of the DFT/MRCI(2) vertical excitation
      energies from those computed from the diagonalization of the
      DFT/MRCI Hamiltonian for the molecules in Thiel's test set.}
    \label{fig:deltae}
  \end{center}
\end{figure}

\begin{figure*}
  \begin{center}
    \includegraphics[width=0.45\textwidth,angle=0]{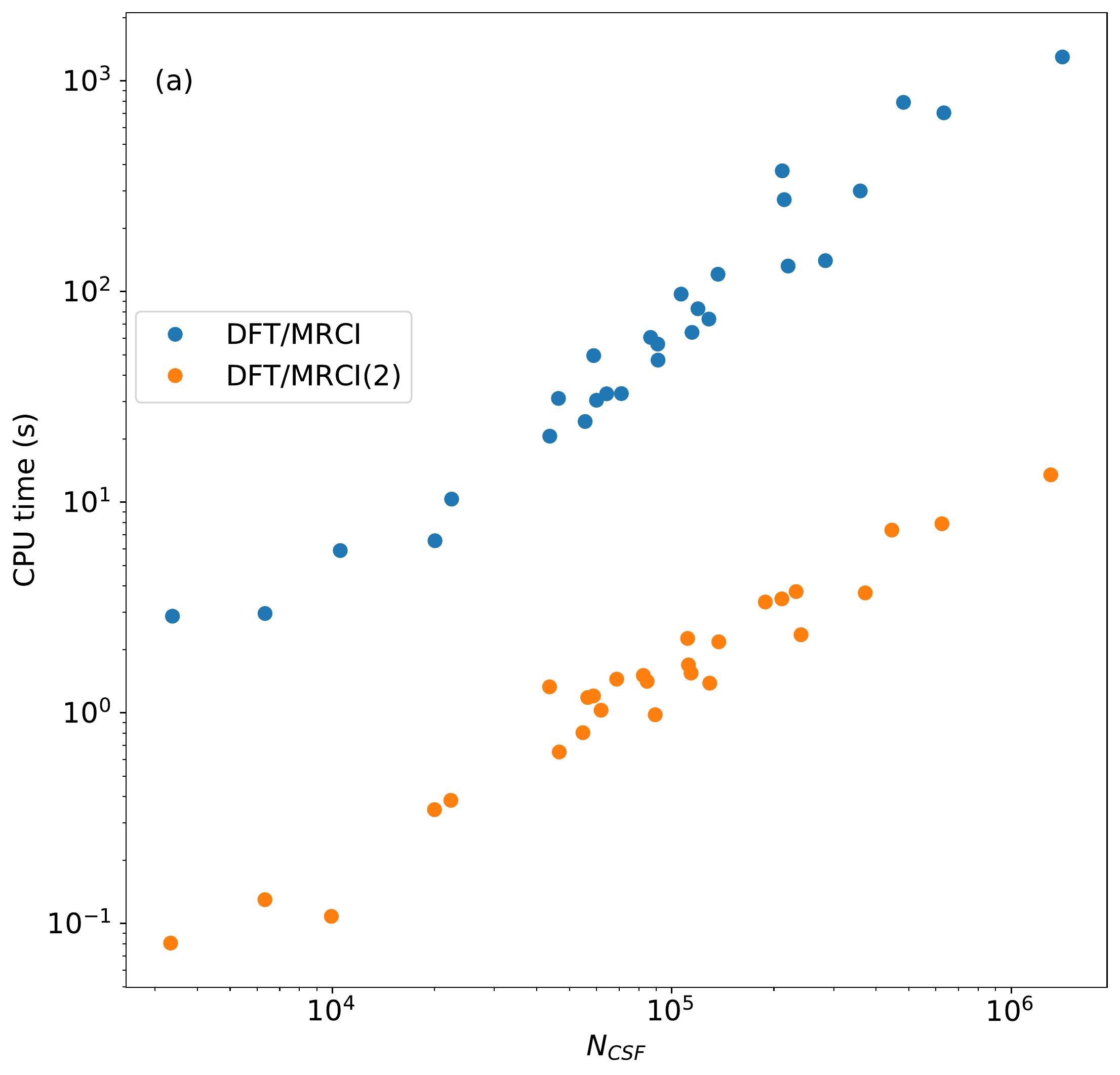}
    \includegraphics[width=0.45\textwidth,angle=0]{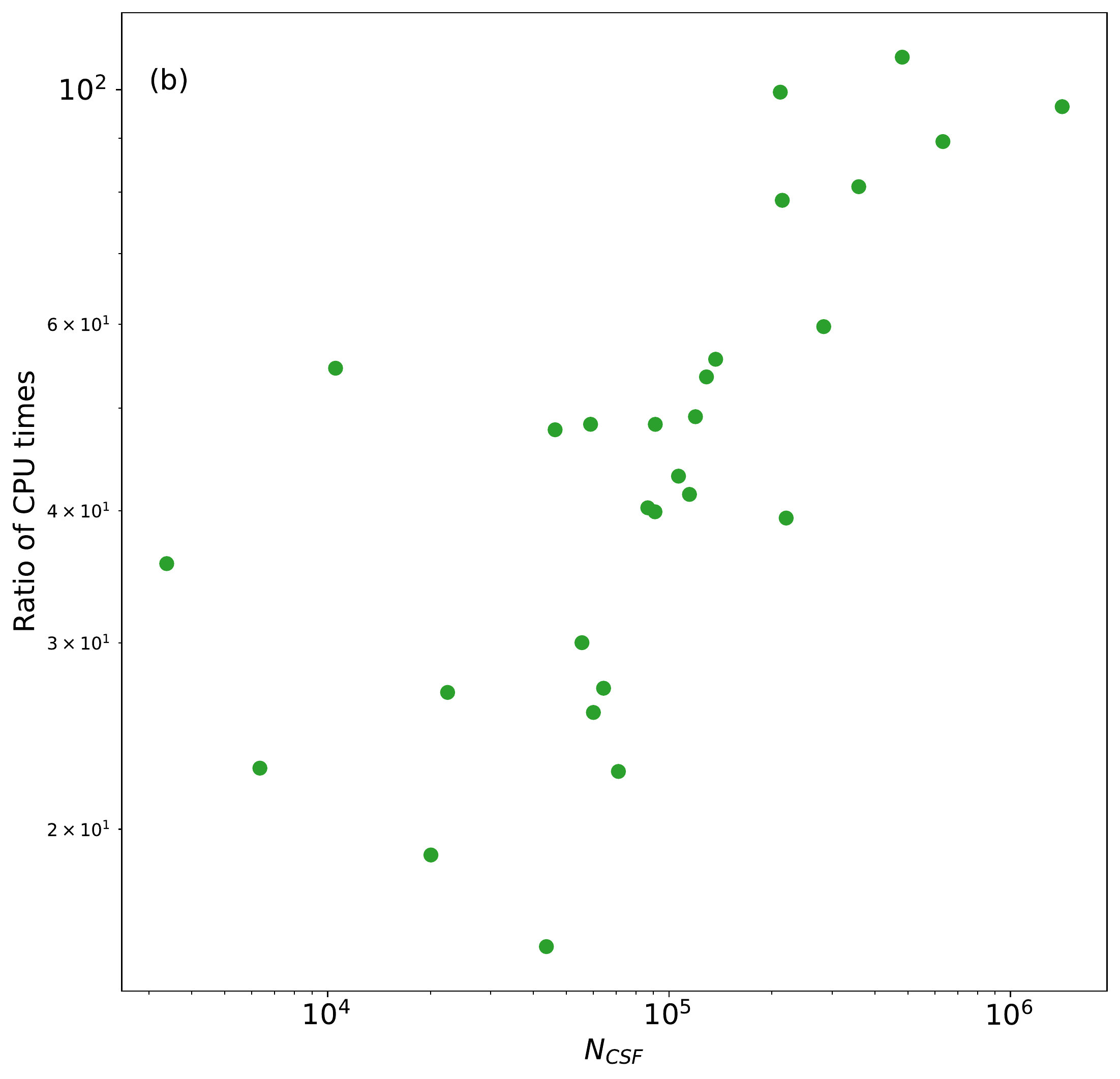}
    \caption{(a) Wall times for the diagonalization of the DFT/MRCI
      Hamiltonian matrix and DFT/MRCI(2) effective Hamiltonian matrix
      as a function of the total number of CSFs. (b) Ratio of the
      DFT/MRCI and DFT/MRCI(2) Hamiltonian diagonalization wall times
      as a function of the total number of CSFs. All calculations were
      performed using a single thread of an Intel Xeon Gold 6130 CPU @
      2.10 GHz.}
    \label{fig:times}
  \end{center}
\end{figure*}

To determine the accuracy of the DFT/MRCI(2) approximation, we
consider the deviation of the eigenvalues $E_{I}^{[2]}$ of the
effective Hamiltonian $\boldsymbol{\mathcal{H}}^{[2]}$ from the
DFT/MRCI energies. To do so, vertical excitation energies were
computed for the members of Thiel's test set of 28 small-to-medium
sized organic molecules\cite{schreiber_2008}. For each molecule in the
set, four singly excited states per irrep were computed, resulting in
a total of 472 excitation energies and accounting for states of both
valence and Rydberg character. All calculations were performed using
the aug-cc-pVDZ basis\cite{dunning_1989} and aug-cc-pVDZ-jkfit
auxiliary basis.

Shown in Figure~\ref{fig:deltae} are the differences, $\Delta E$,
between the excitation energies obtained by diagonalizing the DFT/MRCI
and effective DFT/MRCI(2) Hamiltonians. In general, excellent
agreement is obtained, with a root mean squared deviation (RMSD) of
0.027 eV and a maximum absolute error of 0.138 eV being found for the
472 excited states considered. We also note that 99\% of the
DFT/MRCI(2) excitation energies are within 0.1 eV of the DFT/MRCI
values. This value falls to 94\% for an error bound of 0.05 eV. We
thus conclude that DFT/MRCI(2) is, in general, an excellent
approximation to the DFT/MRCI method.

As well as offering high precision (in comparison to the parent
DFT/MRCI method), the DFT/MRCI(2) approximation also results in large
computational savings by replacing the iterative diagonalization of
the $N_{CSF} \times N_{CSF}$ DFT/MRCI Hamiltonian with the
construction and diagonalization of the $N_{state} \times N_{state}$
DFT/MRCI(2) effective Hamiltonian
$\boldsymbol{\mathcal{H}}_{DFT}^{[2]}$. Shown in
Figure~\ref{fig:times}~(a) is a comparison of the CPU times for these
steps for all molecules in the test set, shown as a function of the
total number (summed across all irreps) of CSFs. We clearly see that
the replacement of the DFT/MRCI Hamiltonian with the DFT/MRCI(2)
effective Hamiltonian results in significant computational gains for
all molecule in the test set. Indeed, even for the largest system
considered (napthalene, 32 excited states), the construction and
diagonalization of the effective Hamiltonian
$\boldsymbol{\mathcal{H}}_{DFT}^{[2]}$ only requires around 10 seconds
of CPU time. Note that this value accounts for \textit{all} iterations
of the reference space refinement. It is also insightful to consider
the ratio of the CPU times for the diagonalization of the DFT/MRCI
Hamiltonian and DFT/MRCI(2) effective Hamiltonian matrices. These
values are shown plotted in Figure~\ref{fig:times}~(b) as a function
of the total number of CSFs. The computational gains afforded by the
DFT/MRCI(2) approximation are clearly seen, with speedups of around
80-100x being found for the largest molecules in the test set. It is
noteworthy that the ratio of the timings of DFT/MRCI Hamiltonian and
DFT/MRCI(2) effective Hamiltonian diagonalization steps grows with the
size of the CSF basis, a result of the more lower scaling of the cost
of effective Hamiltonian construction with the number of CSFs. We thus
expect that even greater computational savings will be found as the
size of the CSF basis grows.

Furthermore, it is important to note that for large CSF bases
($N_{CSF} > \sim 10^{6}$), a direct, configuration-driven algorithm
becomes necessary in the iterative diagonalization of the DFT/MRCI
Hamiltonian matrix, requiring a Hamiltonian build for every
iteration. In contrast, the DFT/MRCI(2) effective Hamiltonian needs
only to be constructed once in order to compute it's eigenvalues and
the first-order corrected wave functions. Thus, in such situations,
one needs to multiply the ratio of the DFT/MRCI Hamiltonian and
DFT/MRCI(2) effective Hamiltonian build times by the number of
iterations required in the DFT/MRCI Hamiltonian diagonalization step
to arrive at an estimate of the true computational savings afforded by
the DFT/MRCI(2) approximation. As such, we predict that the
computational savings afforded by the DFT/MRCI(2) approximation will
reach up to three orders of magnitude when considering large systems.

\subsection{Comparison to p-DFT/MRCI}
Recently, a new approximation to DFT/MRCI, termed p-DFT/MRCI, was
introduced\cite{neville_p-dftmrci}. The idea underlying p-DFT/MRCI is
to reduce the size of the CSF basis following the energy-based
configuration selection by discarding superfluous configurations, as
determined by a perturbative estimate of their contribution to the
DFT/MRCI wave functions of interest. The DFT/MRCI Hamiltonian matrix
is then built and diagonalized within the space spanned by the
resulting pruned CSF basis, with the energetic contribution from the
discarded CSFs being accounted for perturbatively. Considering their
reliance on perturbation theories, it is worth while considering the
relative performance and strengths of the p-DFT/MRCI and DFT/MRCI(2)
methods.

First, we note that both methods use the Epstein-Nesbet partitioning
of the Hamiltonian. In the case of DFT/MRCI(2), this is combined with
GVVPT2 to arrive at an effective Hamiltonian formalism. In the
p-DFT/MRCI approach, second-order RSPT is used in both the CSF pruning
and perturbative energy correction steps (see
Reference~\citenum{neville_p-dftmrci} for details). In both cases,
however, the key quantities required to be computed are the B-vectors

\begin{equation}
  \boldsymbol{B}_{I} = \langle \boldsymbol{\Omega} | \hat{H}^{DFT} |
  \Psi_{I}^{(0)} \rangle.
\end{equation}

\noindent
In p-DFT/MRCI, these are computed in, and form the bottleneck of, the
pruning step, and are re-used in the application of the perturbative
energy corrections. The calculation of the B-vectors is \textit{the}
bottleneck in a DFT/MRCI(2) calculation, being required to compute the
effective Hamiltonian (see Equation \ref{eq:dftmrci2_ham}). Thus, the
calculation of the DFT/MRCI(2) effective Hamiltonian has essentially
the same cost as the pruning step alone in a p-DFT/MRCI
calculation. As such, a DFT/MRCI(2) calculation should always be
computationally cheaper than the corresponding p-DFT/MRCI one.

\begin{figure}
  \begin{center}
    \includegraphics[width=0.45\textwidth,angle=0]{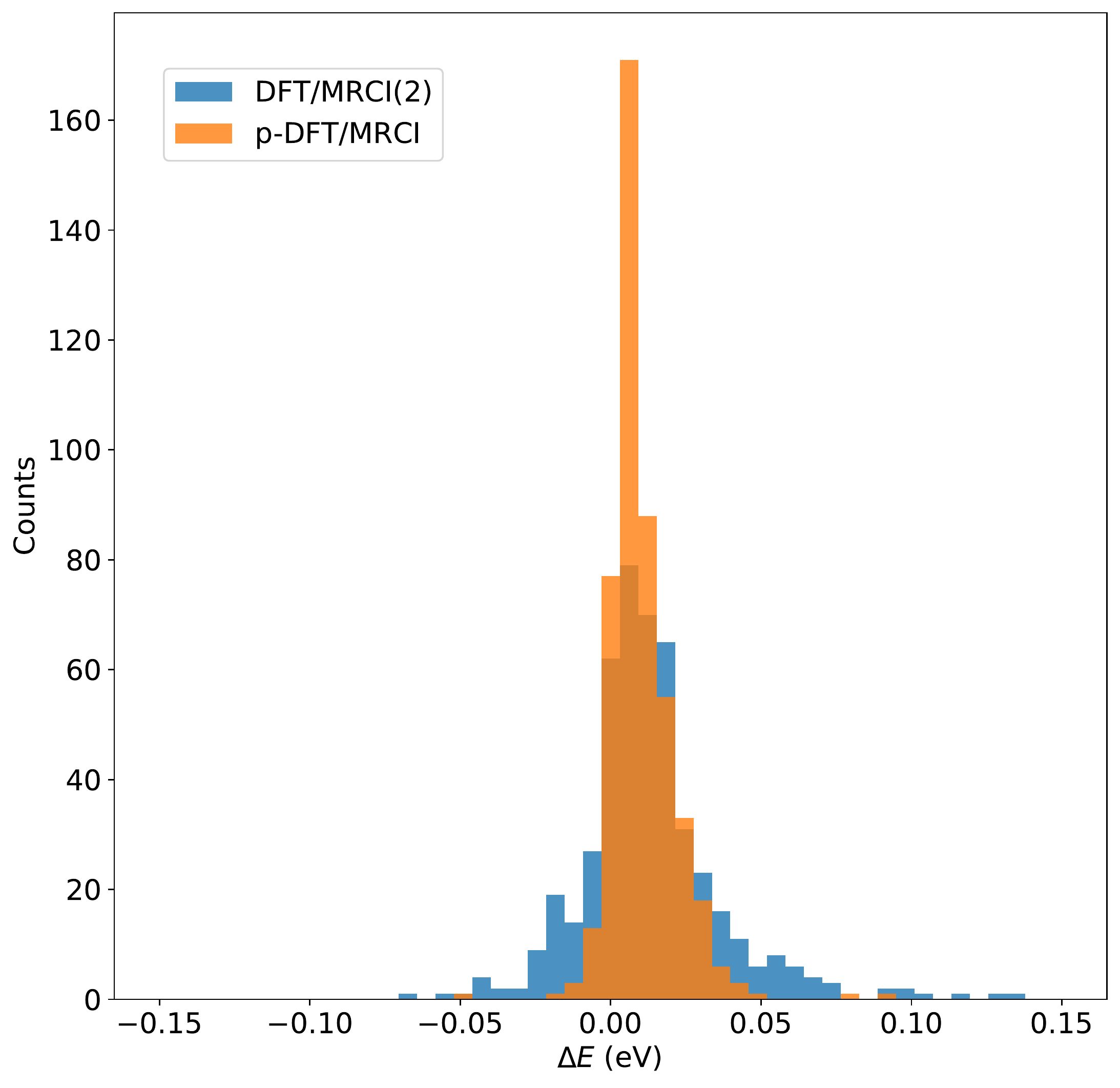}
    \caption{Differences of the DFT/MRCI(2) and p-DFT/MRCI vertical
      excitation energies from those computed from the diagonalization
      of the DFT/MRCI Hamiltonian for the molecules in Thiel's test
      set. In the p-DFT/MRCI calculations, a pruning threshold of
      $\alpha_{p}=0.90$ was used.}
    \label{fig:deltae_comp}
  \end{center}
\end{figure}

We next consider the relative accuracies of DFT/MRCI(2) and p-DFT/MRCI
relative to the original DFT/MRCI method. Shown in
Figure~\ref{fig:deltae_comp} are the deviations $\Delta E$ of the
DFT/MRCI(2) and p-DFT/MRCI excitation energies from the DFT/MRCI
values for the molecules in Thiel's test set. In the p-DFT/MRCI
calculations, a pruning threshold of $\alpha_{p}=0.90$ was used (see
Reference~\citenum{neville_p-dftmrci} for a definition of this
quantity). It is found that DFT/MRCI(2) yields excitation errors of
lower accuracy than p-DFT/MRCI. However, this deterioration of
accuracy is slight, with the RMSDs increasing from 0.015 to 0.027
eV. Furthermore, similar maximum absolute errors of 0.086 and 0.138 eV
are found for p-DFT/MRCI and DFT/MRCI(2), respectively. We thus
conclude that DFT/MRCI(2) offers a very similar level of accuracy as
p-DFT/MRCI but at reduced computational cost.

\subsection{Example calculations}

\begin{figure*}
  \begin{center}
    \includegraphics[width=1.0\textwidth,angle=0]{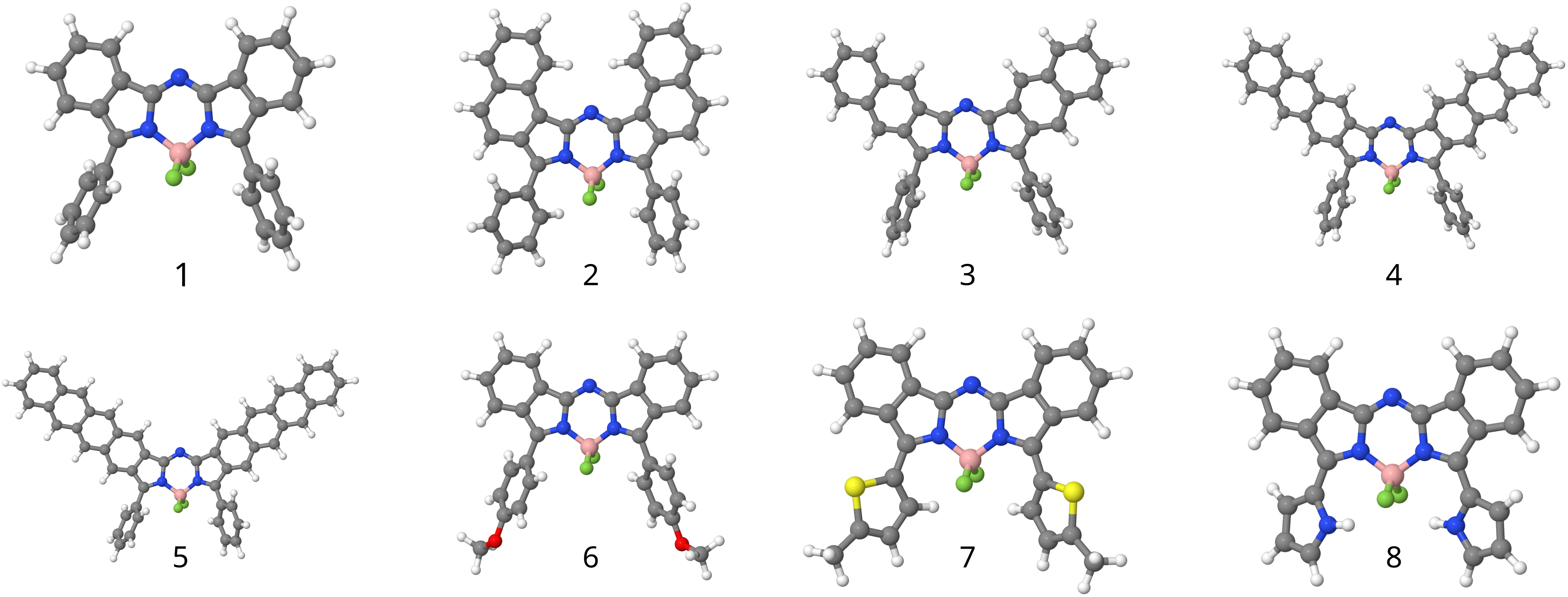}
    \caption{Set of aza-BODIPY molecules and numbering used in the
      example DFT/MRCI(2) calculations.}
    \label{fig:bodipy_geoms}
  \end{center}
\end{figure*}

Having established the accuracy of the DFT/MRCI(2) approximation
relative to the parent DFT/MRCI method, we now consider its
application to a problem of real world interest: the $S_{1}$ vertical
excitation energies of a range of aza-boron-dipyrromethane
(aza-BODIPY) derivatives. This class of molecules are the subject of
intense study given their use as near-infrared dyes in applications
ranging from photovoltaics to bioimaging and photodynamic
therapy\cite{erten-ela2008, rousseau2009, whited2011, flavin2011,
  iehl2012, zheng2008, bouit2009, didier2009, coskun2007, murtagh2009,
  han2009, shi2012, killroan2002, adarsh2010}. The nine aza-BODIPY
molecules shown in Figure~\ref{fig:bodipy_geoms} were chosen as
representative examples. A recent study of these molecules by
Berraud-Pache \textit{et al.}\cite{berraud-pache_2019} provides
$S_{1}$ vertical excitation energies computed at the similarity
transformed equation of motion coupled cluster with singles and
doubles level of theory within the domain-based pair natural orbital
approximation (DLPNO-STEOM-CCSD)\cite{nooijen_steomccsd1997,
  riplinger2013I, riplinger2013II, dutta2017, dutta2018}. This allows
us to make a meaningful comparison to excitation energies calculated
at a near-benchmark \textit{ab initio} level of theory.

As in Berraud-Pache \textit{et al.}'s DLPNO-STEOM-CCSD calculations,
the DFT/MRCI(2) calculations were performed using the def2-TZVP(-f)
basis, along with the def2-universal-JKFIT auxiliary basis. In all
calculations, an energy-based configuration selection threshold of
$\delta E_{sel}=1.0$ E$_{\text{h}}$ was used. The DLPNO-STEOM-CCSD
calculations of Berraud-Pache \textit{et al.} were performed using an
implicit solvation model to simulate solvation in dicholoromethane. To
account for this, the DFT/MRCI(2) calculations were performed in
conjunction with the domain-decomposed COSMO (ddCOSMO) solvation
model\cite{cances2013, lipparini2013}, using a dielectric constant of
8.93. The aza-BODIPY ground state minimum geometries were taken from
Reference~\citenum{berraud-pache_2019}.

\begin{table*}
  \centering
  \caption{$S_{1}$ vertical excitation energies, in units of eV, for
    the aza-BODIPY molecules calculated at the DFT/MRCI(2) and
    DLPNO-STEOM-CCSD levels of theory using def2-TZVP(-f)
    basis. DLPNO-STEOM-CCSD values are taken from
    Reference~\citenum{berraud-pache_2019}. A configuration selection
    threshold of $\delta E_{sel} = 1.0$ E$_{\text{h}}$ was used in all
    DFT/MRCI(2) calculations. Also given are the excitation energy
    differences, $\Delta E$, between the two levels of theory as well
    as the weight $| \langle \overline{\text{w}} \omega | \Psi_{0}
    \rangle |^{2}$ of the base configuration $\overline{\text{w}}$ in
    the DFT/MRCI(2) ground state wave functions.}
  \begin{threeparttable}
    \begin{tabular}{ccccc}
      \hline
      Molecule & DFT/MRCI(2) & DLPNO-STEOM-CCSD & $\Delta E$ & $| \langle \overline{\text{w}} \omega | \Psi_{0} \rangle |^{2}$ \\
      \hline
      1 & 1.79 & 1.72 & -0.07 & 0.88 \\
      2 & 1.63 & 1.60 &  0.03 & 0.87 \\
      3 & 1.40 & 1.44 &  0.04 & 0.84 \\
      4 & 1.10 & 1.23 & -0.13 & 0.82 \\
      5 & 0.87 & 1.09 & -0.22 & 0.80 \\
      6 & 1.73 & 1.64 &  0.09 & 0.88 \\
      7 & 1.61 & 1.55 &  0.06 & 0.88 \\
      8 & 1.59 & 1.58 &  0.01 & 0.87 \\
      9 & 1.61 & 1.52 &  0.09 & 0.88 \\
      \hline
    \end{tabular}
  \end{threeparttable}
  \label{table:bodipy_vees}
\end{table*}

\begin{table}
  \centering
  \caption{CSF basis dimensions, $N_{CSF}$, and CPU times (in units of
    seconds) for the aza-BODIPY DFT/MRCI(2) calculations. In all
    calculations, the def2-TZVP(-f) basis was used along with an
    energy-based configuration selection threshold of $\delta E_{sel}
    = 1.0$ E$_{\text{h}}$. All CPU times correspond to the use of
    a single thread of an Intel Xeon Gold 6130 CPU @ 2.10 GHz.}
  \begin{threeparttable}
    \begin{tabular}{ccc}
      \hline
      Molecule & $N_{CSF}$ / $10^{6}$ & CPU time \\
      \hline
      1 & 2.4  & 72  \\
      2 & 6.9  & 170 \\
      3 & 7.1  & 142 \\
      4 & 17.5 & 308 \\
      5 & 42.7 & 636 \\
      6 & 3.5  & 99  \\
      7 & 2.7  & 86  \\
      8 & 1.3  & 53  \\
      9 & 1.5  & 49  \\
      \hline
    \end{tabular}
  \end{threeparttable}
  \label{table:bodipy_times}
\end{table}

Shown in Table~\ref{table:bodipy_vees} are the DFT/MRCI(2) $S_{1}$
vertical excitation energies alongside the reference DLPNO-STEOM-CCSD
values for all nine aza-BODIPY molecules. In general, the agreement
with the DLPNO-STEOM-CCSD results is excellent, with an RMSD of 0.10
eV being found. The largest deviations, -0.13 and -0.22 eV, occur for
molecules 4 and 5, respectively. Interestingly, for these molecules,
the DFT/MRCI(2) ground state wave function is found to contain
significant contributions from non-base configurations: those with
non-HF occupations. We find a significant contribution from multiple
doubly-excited configurations, in particular those corresponding to
excitation from the HOMO to LUMO KS MOs. As a measure of this, we also
give in Table~\ref{table:bodipy_vees} the weights $|\langle
\overline{\text{w}} \omega | \Psi_{0} \rangle|^{2}$ of the base
configuration $\overline{\text{w}}$ for the DFT/MRCI(2) ground state
$|\Psi_{0}\rangle$ of each molecule. This base configuration weight is
smallest for molecules 4 and 5, taking values of 0.84 and 0.81,
respectively. This result is in line with the those of Momeni and
Brown\cite{brown_bodipy2015}, in which the ground state wave functions
of a large number of BODIPY molecules were found to contain
significant contributions from doubly-excited configurations at the
CASSCF level of theory. This suggests that a single reference method
such as DLPNO-STEOM-CCSD \textit{might} struggle to give a correct
description of these systems, and may explain the larger discrepancy
between the DFT/MRCI(2) and DLPNO-STEOM-CCSD excitation energies for
these two molecules. A thorough investigation of this is, however,
beyond the scope of the current paper, but may constitute an
interesting avenue of future research.

Finally, we consider the computational efficiency of the DFT/MRCI(2)
calculations. Although absolute timings are limited in meaning, being
both hardware and implementation dependent, the CPU times for the
example aza-BODIPY calculations do provide some insight into the high
efficiency of the DFT/MRCI(2) method. These are shown, in
Table~\ref{table:bodipy_times} alongside the size of the final,
converged CSF bases generated by the iteratively refined reference
space. We note that these calculations were performed in serial on a
single thread of an Intel Xeon Gold 6130 CPU @ 2.10 GHz. For the
smallest calculations, molecules 8 and 9, with CSF basis sizes of
$1.3\times10^{6}$ and $1.5\times 10^{6}$, respectively, the cumulative
cost of all steps involved in the DFT/MRCI(2) calculations took less
one minute. In the largest calculation (molecule 5), the dynamic
configuration selection algorithm (see Equation~\ref{eq:dynsel})
generated an extremely large CSF basis of dimension
$42.7\times10^{6}$. However, the DFT/MRCI(2) calculation remained
eminently tractable, requiring just over ten minutes in total.

\section{Conclusions and outlook}\label{sec:conclusions}
We have presented a new perturbative approximation to the DFT/MRCI
method based on a combination of quasi-degenerate perturbation theory
and the Epstein-Nesbet Hamiltonian partitioning. The resulting method,
termed DFT/MRCI(2), replaces the iterative diagonalization of the
DFT/MRCI Hamiltonian with the construction and full diagonalization of
a small effective Hamiltonian matrix determined up to second-order in
perturbation theory. This replacement of the full Hamiltonian
diagonalization step results in orders of magnitude savings in terms
of computational costs. Moreover, the resultant errors in the computed
excitation errors are small, with an RMSD of 0.027 eV being found for
an extensive set of excited states of the molecules in Thiel's test
set\cite{schreiber_2008}.

In terms of potential applications, DFT/MRCI(2) seems ideally suited
to situations in which either: (i) large numbers of excited states are
required to be computed, and/or; (ii) potential energies (and
potentially non-adiabatic couplings) are required at a large number of
nuclear geometries. Scenario (i) is routinely encountered, e.g., in
the calculation of X-ray absorption spectra for large molecules, where
the number of core-excited states needed to be computed may number in
the hundreds. Here, the DFT/MRCI(2) approximation will be advantageous
due to: (a) a massive reduction in the cost associated with the
(effective) Hamiltonian construction, as well as; (b) the complete
obviation of the need to orthogonalize large numbers of wave
function-sized vectors against each other that occurs in iterative
diagonalization schemes. Scenario (ii) will occur if one is interested
in the use of DFT/MRCI in quantum dynamics simulations. Here, the
sheer reduction in computation cost afforded by the diagonalization of
the DFT/MRCI(2) effective Hamiltonian instead the DFT/MRCI Hamiltonian
imbues the method with much promise. Additionally, it will be trivial
to pair DFT/MRCI(2) with a propagative block diagonalization
diabatization scheme such as that detailed in
Reference~\citenum{ours_dftmrci_pbdd}, allowing access to
non-adiabatic coupling terms. As such, DFT/MRCI(2) offers much promise
in the area of on-the-fly non-adiabatic quantum dynamics
simulations. Both these areas of applications will be the subject of
future work within our group.

%\bibliography{./ref}

\end{document}